\documentclass{iopart}
\usepackage{graphicx,times,iopams,color}

\newcommand{\abs}[1]{\left|#1\right|}

\newcommand{\ket}[1]{|#1\rangle}

\newcommand{\boket}[3]{\langle\, #1 \,|\, #2 \,|\, #3 \,\rangle}

\newcommand{\be}{\begin{equation}}
\newcommand{\ee}{\end{equation}}
\newcommand{\bc}{\begin{center}}
\newcommand{\ec}{\end{center}}

\newcommand{\eqref}[1]{(\ref{#1})}

\begin{document}

\title[Dispersive regime of the Jaynes-Cummings and Rabi lattice]{Dispersive regime of the Jaynes-Cummings and Rabi lattice}
\author{Guanyu Zhu$^1$, Sebastian Schmidt$^2$ and Jens Koch$^1$}
\address{$^1$ Department of Physics and Astronomy, Northwestern University, Evanston, IL~60208, USA}
\address{$^2$ Institute of Theoretical Physics, ETH Zurich, 8093 Zurich, Switzerland}
\ead{jens-koch@northwestern.edu}

\begin{abstract} 
Photon-based strongly-correlated lattice models like the Jaynes-Cummings and Rabi lattices differ from their more conventional relatives like the Bose-Hubbard model by the presence of an additional tunable parameter: the frequency detuning between the pseudo-spin degree of freedom and the harmonic mode frequency on each site. Whenever this detuning is large compared to relevant coupling strengths, the system is said to be in the dispersive regime. The physics of this regime is  well-understood at the level of a single Jaynes-Cummings or Rabi site. Here, we extend the theoretical description of the dispersive regime to lattices with many sites, for both strong and ultra-strong coupling. We discuss the nature and spatial range of the resulting qubit-qubit and photon-photon coupling, demonstrate the emergence of photon-pairing and squeezing, and illustrate our results by exact diagonalization of the Rabi dimer.
\end{abstract}


\submitto{\NJP}


\section{Introduction}

Bosonic and fermionic lattice systems, realized experimentally with ultracold atoms or trapped ions, have long served as a paradigm for quantum simulators of strongly-correlated many-body systems \cite{Lewenstein2007,Bloch2008,Barreiro2012}. Recently, Jaynes-Cummings and Rabi lattices have been suggested as an exciting variant of this idea \cite{Hartmann2006,Greentree2006,Angelakis2007}. The basic physics of these lattices may enable new types of photon-based quantum simulators suitable for exploring many-body physics in and out of equilibrium, and has stirred lively interest \cite{Hartmann2008,Tomadin2010,Houck2012,Schmidt:2012us}.

 In the Jaynes-Cummings and Rabi lattice, each lattice site consists of a photon mode interacting locally with a two-level system, and is described by the ordinary Jaynes-Cummings \cite{Jaynes1963} or Rabi model  \cite{Rabi:1936vf}. In addition, photons are allowed to hop between nearest-neighbor lattice sites.  A key difference between the traditional boson and fermion lattices and the less conventional Jaynes-Cummings and Rabi lattice is that the latter provide an interesting additional tunable parameter: the detuning $\Delta$. Its origin lies in the two-component nature of the Jaynes-Cummings and Rabi model, comprising an electromagnetic field component  and a matter component. 
Each component is associated with its own characteristic energy scale. The possible energy mismatch between the two is quantified by the detuning $\Delta$.

By changing the detuning, three qualitatively different regimes of the Jaynes-Cummings lattice and Rabi lattice can be accessed. In the resonant regime, the detuning is small and photons and matter excitations readily hybridize to form polaritons. In the dispersive regime, the magnitude $\abs{\Delta}$ of the detuning  between the two-level splitting $\epsilon$ and the photon frequency $\omega$  is large. According to the sign of $\Delta=\epsilon-\omega$, we distinguish between  the dispersive regime with negative detuning, where low-energy physics predominantly  involves matter excitations, and the dispersive regime with positive detuning, in which photons govern the behavior at low energies.  In this paper, we focus primarily on the two dispersive regimes where all interaction strengths are small compared to $\abs{\Delta}$. 

The circuit QED architecture \cite{Blais2004,Wallraff2004,Schoelkopf2008} constitutes one of the most promising experimental platforms for the realization of dispersive Jaynes-Cummings and Rabi lattice systems \cite{Houck2012, Schmidt:2012us}. In circuit QED lattices, photons can hop between transmission line resonators (which are coupled capacitively) and locally interact with a superconducting qubit. The qubit energy $\epsilon$ (and hence the detuning parameter $\Delta$) can be tuned in-situ by an externally applied magnetic flux.  Recently, the coherent photon exchange via hopping between several coupled resonators has also been realized experimentally \cite{Underwood2012}.

Theoretically, the dispersive regime for a single Jaynes-Cummings \cite{Blais2004} or Rabi system \cite{Zueco2009} is well understood. Its usual description is based on employing a perturbative Schrieffer-Wolff transformation, which eliminates the Jaynes-Cummings or Rabi coupling by switching to an appropriate dressed-state basis. 

Here, we extend this procedure to an entire lattice of sites and systematically discuss all contributions in second-order perturbation theory (section \ref{sec2}).  
We find that effective qubit-qubit interactions\footnote{The term `qubit' will serve as a shorthand for `two-level system' in the following.}, qubit-state dependent photon hopping terms and photon pairing terms (squeezing terms) emerge. All inter-site interactions are short-range but not limited to nearest-neighbor sites.  We explore the implications of the derived effective Hamiltonian for the low-energy physics of the Jaynes-Cummings and Rabi lattices in section \ref{sec:phrab}. For negative detuning, we show that the system reduces to an effective spin model with  XY-type interaction for the Jaynes-Cummings lattice, and to a transverse Ising model for the Rabi lattice. For positive detuning, we discuss in detail the one-mode and two-mode vacuum squeezing relevant in the ultra-strong coupling regime as described by the dispersive Rabi model.   Due to the  ultra-strong coupling, approached indeed in a recent experiment for a single site  in circuit-QED architecture \cite{Niemczyk2010}, the non-trivial nature of the ground state makes the Rabi lattice particularly interesting.   In section \ref{sec4}, we  study the application of the dispersive regime to the Rabi dimer and confirm the validity of the derived effective Hamiltonians by comparison with results from exact diagonalization.   Finally, we summarize and give an outlook on questions of future interest in section \ref{sec5}.

\section{Derivation of the effective Hamiltonian\label{sec2}}
The Rabi lattice is described by the model Hamiltonian
\be\fl\label{Rlat}
H= \omega  \sum_j  a^\dag_j a_j + \epsilon \sum_j  \sigma^+_j   \sigma^-_j + g \sum_j \left( a_j \sigma^+_j + a_j^\dag \sigma^+_j  +  \mathrm{H.c.} \right) +  t\sum_{\langle j,j' \rangle}\left(a_j^\dag a_{j'}  + \mathrm{H.c.} \right).
\ee
Here, each Rabi site is labeled by an index $j$, and consists of a harmonic mode coupled to a pseudo spin-$1/2$. As usual, excitations of the harmonic mode and spin on site $j$ are created by $a_j^\dag$ and $\sigma_j^+$, and annihilated by their Hermitean-conjugate counterparts. The energy necessary for an excitation of either type is fixed by $\omega$ and $\epsilon$, respectively. The coupling strength is set by the parameter $g$. From the beginning on, we include the counter-rotating  terms characteristic of the Rabi model and the ultra-strong coupling regime, and only drop those terms in our discussion of the Jaynes-Cummings limit where the rotating-wave approximation (RWA) is applicable (see text below). For the entirety of this paper, we will focus on the disorder-less case, assuming that all lattice sites have identical parameters. In the following and for the sake of brevity, we simply refer to the pseudo-spin degree of freedom as ``qubit''. 

In addition to the onsite Rabi Hamiltonians, the last term in equation \eqref{Rlat} introduces coupling between resonators. We consider weak resonator coupling, $|t| \ll \omega, \epsilon$, and treat it within the RWA. As written in equation \eqref{Rlat}, the coupling term then allows for photons to hop from one resonator to its nearest neighbors, as indicated by the angular brackets in the summation over resonator pairs. The sign of the hopping amplitude $t$ depends on the specific system realization. In the circuit QED architecture, for example, it depends on whether full-wavelength modes ($t>0$) or half-wavelength modes ($t<0$) are used \cite{Nunnenkamp2011,Schmidt:2012us}.  

The Rabi lattice \eqref{Rlat} enters the dispersive regime when the detuning $\abs{\Delta}=\abs{ \epsilon -\omega}$ between qubit and resonator frequency is large compared to the coupling strength $g$. While the sum frequency $\Sigma=\epsilon+\omega\ge|\Delta|$ obeys $\Sigma\gg|\Delta|$ whenever the RWA holds, a hallmark of the dispersive regime of the Rabi model beyond RWA is that detuning and frequency sum may be of the same order, $\Sigma\sim|\Delta|$ \cite{Zueco2009}. In this case, two sub-regimes are of particular interest: the dispersive Rabi regime with negative detuning, $\epsilon\ll g\ll \omega$, and the dispersive Rabi regime with positive detuning, $\omega \ll g \ll \epsilon$. Both sub-regimes will be investigated in section \ref{sec:phrab}.

In any of the mentioned cases, the idea underlying the dispersive regime remains the suppression of interconversion between qubit and photon excitations due to a large energy mismatch which is not overcome by the coupling $g$. Under these conditions, the Rabi interaction terms $\sim g$ can be treated perturbatively. 

To obtain a simple effective Hamiltonian describing the dispersive regime, it is convenient to carry out the perturbation theory in the form of a unitary Schrieffer-Wolff transformation, $\tilde{H}=e^{i S} H e^{-i S}$ \cite{Schrieffer1966,Cohen-Tannoudji1998}. For the cases of a single Rabi site and multiple qubits strongly coupled to a single resonator, this procedure was previously discussed by Zueco et al.\ \cite{Zueco2009}. Here, we extend it to a lattice of coupled Rabi sites. Using the appropriate Hermitian generator $S$, the unitary transformation switches to a dressed-state basis in which the original Rabi interaction is eliminated.  Being unitary, the transformation clearly preserves the spectrum of the original Hamiltonian. 

 In our case, the unperturbed Hamiltonian $H_0$ consists of the terms in equation \eqref{Rlat} with the exception of the Rabi coupling term:
\be
H_0
= \omega  \sum_j  a^\dag_j a_j +   t\sum_{\langle j,j' \rangle}\left(a_j^\dag a_{j'}  + \mathrm{H.c.} \right) + \epsilon \sum_j  \sigma^+_j   \sigma^-_j.
\ee
This corresponds to a photonic tight-binding model and uncoupled qubit.
To avoid distractions, we limit our discussion to one-dimensional Rabi lattices. The generalization to lattices with more complex structure is straightforward. Thus, considering a one-dimensional photon tight-binding model with Born-von Karmann periodic boundary conditions over a number $N$ of lattice sites, we define the itinerant photon operators by
\be
\mathsf{a}_k = \frac{1}{\sqrt{N}} \sum_{j=1}^{N} e^{-i k j} a_j, \qquad {\textstyle k=0,\,
\frac{2\pi}{N},\,2\frac{2\pi}{N},\,\ldots,\,(N-1)\frac{2\pi}{N}}.
\ee
Here, for a large number of sites ($N\to\infty$), the quasi-momentum $k$ spans the entire first Brillouin zone, $-\pi<k\le\pi$. Employing this basis, the unperturbed Hamiltonian can be rewritten in the diagonal form as
\be\label{unperturbed}
H_0
= \sum_k \omega_k  \mathsf{a}^\dag_k \mathsf{a}_k+   \epsilon \sum_j  \sigma^+_j   \sigma^-_j.
\ee
Here, the dispersion of the itinerant photons for a one-dimensional lattice is simply $\omega_k= \omega+ 2t \cos{k}$.
 
\begin{figure}[t]
\hspace*{2.5cm}
\includegraphics[width=0.4\columnwidth]{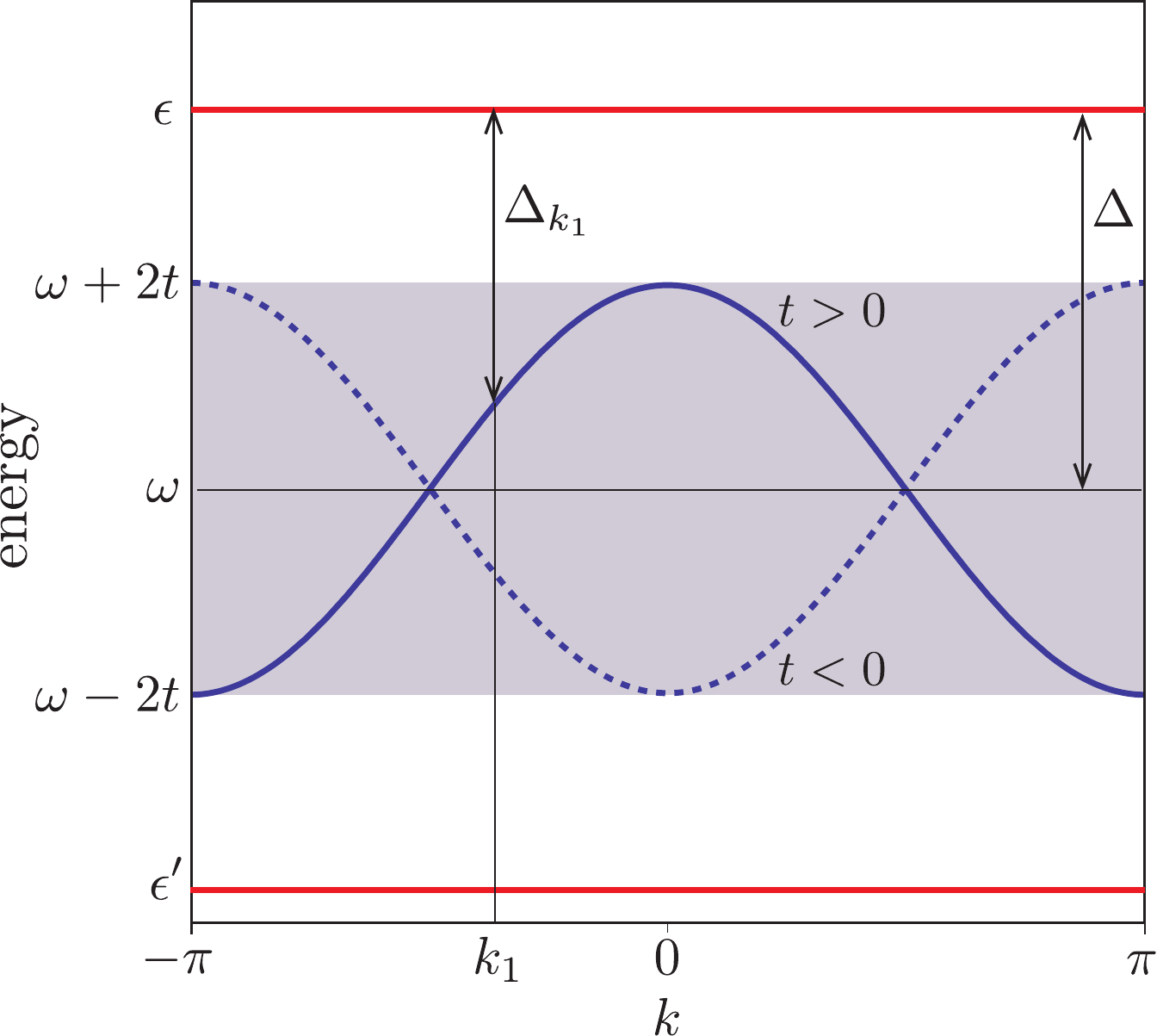}
\caption{Dispersive regime of the Rabi lattice: relevant energy scales.  The plot shows the itinerant photon dispersion $\omega_k$ for positive (blue solid) and negative (blue dashed) hopping amplitude. The shaded region indicates the full width $4t$ of the photon band, centered at the bare photon frequency $\omega$.   Examples for qubit energies with positive ($\epsilon$) and negative ($\epsilon'$) detuning are shown in red.  Expressed in terms of quasi-momenta, the detuning $\Delta_k$ becomes $k$ dependent (see example for $k=k_1$, shown for the $t>0$ case).}
\label{dispersionrelation}
\end{figure}

The perturbation $V$ is  the onsite Rabi interaction on each lattice site and can be expressed in terms of the itinerant photon operators as
\be
V =\frac{g}{\sqrt{N}} \sum_k \sum_j \left( \mathsf{a}_k \sigma^+_j e^{ikj}  + \mathsf{a}_k \sigma^-_j e^{ikj}+  \mathrm{H.c.} \right).
\ee
For itinerant photons, the energy gap towards the qubit energy now depends on the quasi-momentum $k$. We thus define the $k-$dependent  detuning parameter $\Delta_k=\epsilon-\omega_k$ and the frequency sum  $\Sigma_k =\epsilon + \omega_k$.
Ensuring that the qubits remain detuned from \emph{all} itinerant photons, we recognize that 
\be\label{dispersivecondition}
\Delta_\mathrm{min}\equiv \min_k \abs{\Delta_{k}}  = \min_k \abs{\epsilon-\omega_k} \gg g, 
\ee
 constitutes a necessary condition for the dispersive regime of the Rabi lattice model.  Figure \ref{dispersionrelation} illustrates this condition for both positive and negative detuning.

\begin{figure}
\includegraphics[width=1.0\columnwidth]{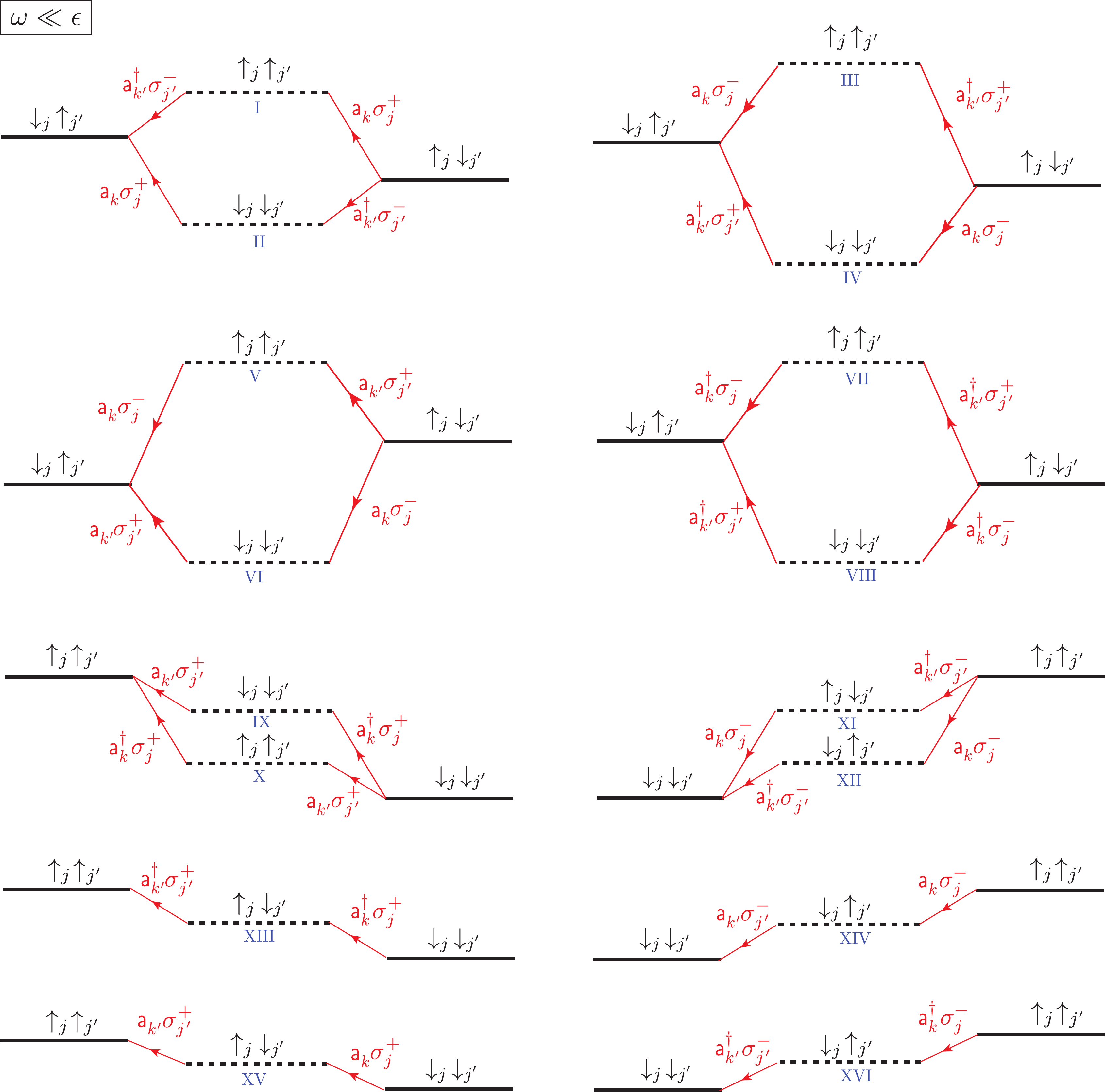}
\caption{``Ladder-type" diagrams used in deriving the effective Hamiltonian.  Horizontal ladder steps represent unperturbed eigenstates of $H_0$ and red arrows the virtual transitions between them. Dashing of horizontal steps indicates virtual intermediate states. Each diagram is to be read from left to right. Labels on horizontal steps show the relevant qubit configurations and labels on arrows the operators producing the transition. Diagrams with two paths lead to partial cancellation in the Hamiltonian. The shown ordering of energies (as indicated by the vertical position of steps in each diagram) refers to the positive detuning case, $\omega \ll \epsilon$. Completely analogous diagrams, differing only in the ordering of energies, apply to the negative detuning case, $\omega \gg \epsilon$. 
The diagrams {\scshape xiii}--{\scshape xvi} vanish when summed over $k,k'$, and hence do not contribute to  the effective Hamiltonian.}
\label{ladders1}
\end{figure}

In the perturbative approach, the Schrieffer-Wolff transformation is carried out to a certain order in the interaction strength $g$. In this paper, we present results for the Rabi lattice model to second order in $g$.  The generator necessary to achieve this must be of first order in $g$ \cite{Cohen-Tannoudji1998}, and is given by
\be\label{generatorRabi}
S_1=\frac{i}{\sqrt{N}}  \sum_k \sum_j \left[ \frac{g}{\Delta_k}\mathsf{a}^\dag_k \sigma^-_j e^{-ikj}   +  \frac{g}{\Sigma_k}\mathsf{a}_k \sigma^-_j e^{ikj}-\mathrm{H.c.}   \right].
\ee
We then construct the effective Hamiltonian by expanding the exponentials in $e^{i S_1} H e^{-i S_1}$   and collecting terms up to second order.\footnote{Recall that $S_2$ does not contribute to the second-order effective Hamiltonian since $[H,S_2]=0$, see e.g., Ref.\ \cite{Cohen-Tannoudji1998}.} 
Due to the form of the interaction $V$, the generator $S_1$ splits into 4 terms,  resulting in $4^2=16$ second-order terms  which contribute to the effective Hamiltonian.  (Note that the first-order term vanishes since $[H,S_1]=0$.) Figure \ref{ladders1} visualizes all 16 contributions in the form of ``ladder-type" diagrams \footnote{Note, that the term ``ladder-type" solely refers to the appearance of these contributions as visualized in Fig.~\ref{ladders1} and should not be confused with other types of Feynman ``ladder" diagrams.}, similar to those previously introduced in Ref.\  \cite{Zhu:2013kf}.

Each diagram can be directly translated into a contribution to the effective Hamiltonian according the following rules:
\begin{enumerate}
\item Each ladder step corresponds to one unperturbed eigenstate of $H_0$, with well-defined number of photons and qubit excitations. Virtual intermediate states are marked by dashing.

\item Diagrams are read from right to left. Arrows show virtual transitions produced by a specific operator term from $S_1$ (label on the arrow), e.g., $\mathsf{a}^\dag_{k'}\sigma^-_{j'}$.  The contribution to the effective Hamiltonian contains the same operator combination (including the operator ordering) as shown by the arrow label. For each occurrence of photon operators, traveling-wave factors should be included according to $\mathsf{a}_k\to \mathsf{a}_k e^{ikj}$ and $\mathsf{a}_k^\dag\to \mathsf{a}^\dag_k e^{-ikj}$.

\item Each contributing diagram involves summation over all intermediate labels $j,j',k,k'$ with a $1/N$ prefactor.

\item The energy coefficient for each contribution is given by
$\frac{g^2}{2}\left[\frac{1}{E_R-E_I} + \frac{1}{E_L-E_I}\right],$
where $E_L$, $E_R$ and $E_I$ denote the bare energies of the left, right and intermediate state respectively. 
\end{enumerate}
As an example, the analytical expressions obtained for the first two diagrams read:
\begin{eqnarray}
\label{eqpathI}
\tilde{H}^{\mathrm{I}} = \frac{1}{N} \sum_{k,k'}  {\sum_{j,j'}}   \frac{g^2}{2} \left[\frac{1}{-\Delta_{k'}}+\frac{1}{-\Delta_k}\right]   e^{i(k'j'-kj)}   \mathsf{a}^\dag_k \mathsf{a}_{k'}  \sigma^-_j  \sigma^+_{j'},\\
\label{eqpathII}
\tilde{H}^{\mathrm{II}} = \frac{1}{N} \sum_{k,k'}  {\sum_{j,j'}}   \frac{g^2}{2} \left[\frac{1}{\Delta_{k'}}+\frac{1}{\Delta_k}\right]   e^{i(k'j'-kj)}  \mathsf{a}_{k'}\mathsf{a}^\dag_k \sigma^+_{j'}  \sigma^-_j.
\end{eqnarray}

\begin{figure}
\flushright
\includegraphics[width=0.9\columnwidth]{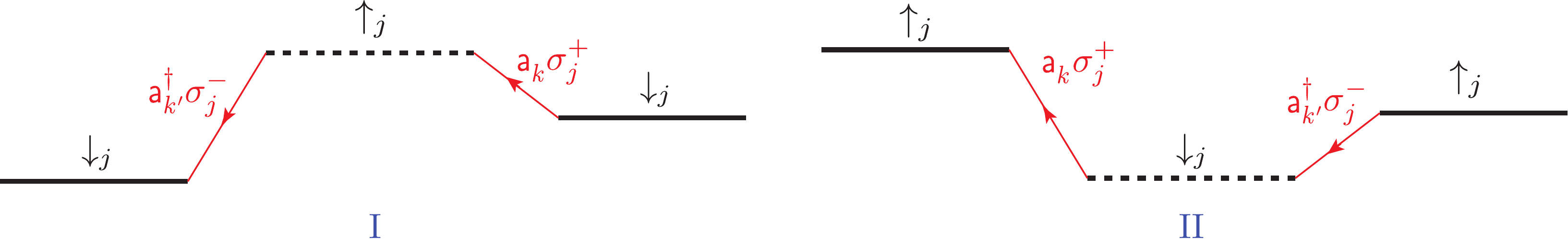}
\caption{Paths from Fig.~\ref{ladders1} for the special case of $j=j'$. As an example, the paths {\scshape i} and {\scshape ii} are shown here. In this case, initial and final states differ,  no interference occurs and photon operators are not cancelled. Such paths contribute to all those second-order terms in the effective Hamiltonian \eqref{Rabieffective} that involve photon operators.}
\label{laddersplit}
\end{figure}

We now turn to the systematic discussion of the contributions obtained from the diagrams shown in figure \ref{ladders1}. All contributions from the paths {\scshape xiii}--{\scshape xvi} vanish. For each of these paths, summation  over $k$ and $k'$ leads to complete cancellation due to the opposite signs of the energy denominators involved. Several paths in Fig.~\ref{ladders1}, including paths {\scshape i} and {\scshape ii}, are shown as pairs. Each member in such a pair has the same initial and final states but undergoes its two virtual transitions in opposite order, thus leading to interference  and partial cancellation. This cancellation originates from the opposite signs of the prefactors, see, e.g., the example in equations \eqref{eqpathI} and \eqref{eqpathII}.  

 We first discuss the case of distinct site indices $j\not=j'$.  Since Pauli operators on different sites commute and itinerant photon operators obey the canonical commutation relation $[\mathsf{a}_k, \mathsf{a}^\dag_{k'}]=\delta_{kk'}$,
we find that all terms in $\tilde{H}^{\mathrm{I}}+\tilde{H}^{\mathrm{II}}$ with $k \neq k'$ and $j\neq j'$ exactly cancel.  In the remaining ($k=k'$) term, all photon operators are eliminated:
\be\label{flfl}
\tilde{H}^{\mathrm{I}} + \tilde{H}^{\mathrm{II}}\bigg|_{j\neq j'} =   \frac{1}{N} \sum_{k}  \sum_{j\not=j'}   \frac{g^2}{\Delta_k}  e^{ik(j'-j)}  \sigma^+_{j'}  \sigma^-_j.
\ee 
Equation \eqref{flfl} represents photon-mediated flip-flop (XY) interaction between qubits on different sites. Further contributions of this type are produced by paths {\scshape iii} and {\scshape iv}. By contrast, the paths {\scshape ix}--{\scshape xii} produce an effective qubit interaction of the type $\sigma^+_{j'}  \sigma^+_j+\sigma^-_{j'}  \sigma^-_j$. Finally, the $j\not=j'$ contributions of the paths {\scshape v}--{\scshape viii} exactly vanish.

Next, we consider the onsite case, i.e., the situation of $j=j'$ in each path. By comparing with the expressions in equations \eqref{eqpathI} and \eqref{eqpathII}, we note that the initial and final states must differ in this special case. For the example of the paths {\scshape i} and {\scshape ii}, the resulting proper diagrams are shown in Fig.~\ref{laddersplit}.
They involve a single qubit flip accompanied by the creation and annihilation of one itinerant photon each.  The $j=j'$ contributions of the paths {\scshape v}--{\scshape viii} produce photon pair creation and annihilation terms.  Finally, the  $j=j'$ contributions of the paths {\scshape ix}--{\scshape xii} identically vanish due to the occurrence of two Pauli raising or lowering operators on the same site, where $({\sigma}^+_j)^2=({\sigma}_j^-)^2=0$.

By transforming all  itinerant photon operators back into real space, we obtain the effective second-order Hamiltonian for the dispersive Rabi regime:

\begin{eqnarray}
\fl  \nonumber \tilde{H}_\mathrm{eff} = H_0 + \frac{g^2}{2}\sum_j \tilde{C}^+_{0} \sigma^z_j
+\frac{g^2}{2}\sum_{j\not=j'} \tilde{C}^{-}_{j - j'} \sigma^x_{j}  \sigma^x_{j'} 
+ g^2 \tilde{C}^+_{0}  \sum_{j} a^\dag_{j}a_{j}\sigma^z_{j}\\\nonumber
\fl\hspace*{1.2cm}
+ \frac{g^2}{2} \tilde{C}^+_{0}  \sum_{j}  (a^\dag_{j}a^\dag_{j}\sigma^z_{j}  + \mathrm{H.c.}) 
+ \frac{g^2}{2} \sum_{j\not=j'} \tilde{C}^+_{j - j'} a^\dag_{j}a_{j'}(\sigma^z_{j}+\sigma^z_{j'})\\
\fl\hspace*{1.2cm}
+ \frac{g^2}{2} \sum_{j<j'} \tilde{C}^+_ {j - j'} (a^\dag_{j}a^\dag_{j'}+a_{j}a_{j'})( \sigma^z_{j}+ \sigma^z_{j'}),
\label{Rabieffective}
\end{eqnarray}
where a global constant has been dropped. The coupling constants $\tilde{C}^{\pm}_m$ depend on the distance $m$ between lattice sites and are defined as 
\be\fl\label{cplf}
\tilde{C}^{\pm}_m= \frac{1}{N} \sum_k \left(\frac{1}{\Delta_k} \pm  \frac{1}{\Sigma_k}\right)e^{imk}=\frac{1}{N\Delta}\sum_k \frac{ e^{imk}}{1-\frac{2t}{\Delta} \cos{k}} \pm \frac{1}{N\Sigma}\sum_k \frac{ e^{imk}}{1-\frac{2t}{\Sigma} \cos{k}}.
\ee

Since the condition \eqref{dispersivecondition} for the dispersive regime of the Rabi lattice also implies that the inequalities $\abs{2t/\Delta}<1$ and $\abs{2t/\Sigma}<1$ must hold, we can ascertain that $g/\Delta$, $g/\Sigma$,  $t/\Delta$ and $t/\Sigma$ are all small parameters. The effective Hamiltonian \eqref{Rabieffective} obtained from the perturbative Schrieffer-Wolff transformation is a series expansion in both $g/\Delta$ and $g/\Sigma$. Due to exact diagonalization of the photon tight-binding model, the coupling constants $\tilde{C}^\pm_m$ contain terms to all orders in $t/\Delta$ and $t/\Sigma$.  
We can further elucidate this fact by re-expressing the denominators in equation \eqref{cplf} in terms of convergent geometrical series, namely
 \be\label{geometric}
\tilde{C}^{\pm}_m=\frac{1}{N}\sum_k e^{imk} \sum_{n=0}^{\infty} \bigg[\frac{1}{\Delta}\left(\frac{t}{\Delta}\right)^n \pm \frac{1}{\Sigma}\left(\frac{t}{\Sigma}\right)^n  \bigg] (e^{ik}+e^{-ik})^n.
\ee
Applying the binomial theorem to the last factor, using the relation $N^{-1}\sum_k e^{ikl}=\sum_{z=-\infty}^\infty\delta_{l,zN}$, and
keeping only the leading order terms in $t/\Delta$ and $t / \Sigma$, we find the approximation 
\be\label{approxcoupling}
\tilde{C}^{\pm}_m \approx \frac{1}{\Delta} \left(\frac{t}{\Delta}\right)^m \pm \frac{1}{\Sigma} \left(\frac{-t}{\Sigma}\right)^m 
\ee
for the coupling constants. Equation \eqref{approxcoupling} is valid for $0\le m \le N/2$. Whenever $m$ is outside this range, the exponents in equation \eqref{approxcoupling} should be replaced by $|m\mathrm{\,mod\,} N|$.
Equations \eqref{Rabieffective}, \eqref{cplf} and \eqref{approxcoupling} constitute the main results of this first part of our paper. In the following we will explore the implications of this effective Hamiltonian and discuss the physics of the dispersive regime of the Rabi lattice and the Jaynes-Cummings lattice in the subsequent section. 

As expected, the dispersive approximation to the Rabi lattice Hamiltonian involves energy shifts and interaction terms which do not inter-convert between qubit and photon excitations.  Based on equation \eqref{Rabieffective}, we now discuss shift and interaction terms one by one. The second term on the right-hand side of equation \eqref{Rabieffective}
captures the Lamb shift for the on-site qubit-resonator system. The shift is identical to the one obtained for a single qubit-resonator system in the dispersive regime \cite{Blais2004} when using the weak-coupling approximation $\tilde{C}^+_0\approx 1/\Delta-1/\Sigma$.

The third term in equation \eqref{Rabieffective} produces photon-mediated qubit-qubit interactions of the  transverse-Ising type.\footnote{Transverse-Ising coupling is also expected for multiple qubits interacting with a single resonator, see Ref.\ \cite{Zueco2009}.} The contributions responsible for going beyond the bare flip-flop (XY) interaction  are the additional counter-rotating terms $\sigma^+_{j} \sigma^+_{j'}$ and $\sigma^-_{j} \sigma^-_{j'}$.
The strength of this interaction is set by the coupling constant $\tilde{C}^-_m$ which, according to equation \eqref{approxcoupling}, decays exponentially with increasing distance $m$ between lattice sites.

The terms 4 -- 7  on the right-hand side of equation (\ref{Rabieffective})  all directly involve photons. Term 4 with the form $a^\dag_j a_{j}\sigma^z_{j}$ produces the well-known AC Stark shift on each lattice site \cite{Blais2004}. Term 5 with the structure $a^\dag_j a^\dag_{j}\sigma^z_{j}$ is an onsite term as well but goes beyond a mere energy shift: here, photon pairs are created or annihilated on a single site. At the same time, the amplitude sign for this process depends on the state of the local qubit. Counter-rotating terms like this one are characteristic of ultra-strong coupling and reflect, naturally, that the total excitation number  $N_\mathrm{tot}=\sum_j (a^\dag_j a_j +  \sigma^+_j   \sigma^-_j )$ is not conserved in the case of the Rabi lattice.

The terms 6 and 7 involve two sites and describe conditional photon hopping and two-mode photon pair creation or annihilation. Remarkably, in both cases the amplitude for these processes depends on the two-qubit operator $(\sigma^z_{j}+\sigma^z_{j'})$ including the $z$-projections of the qubits on the two sites involved in the hopping or pair creation. Assuming qubit configurations composed of $\sigma^z$ eigenstates, hopping and pair creation can be enhanced or suppressed by choosing qubits on the corresponding sites to be aligned or anti-aligned. Again, overall coupling strengths are fixed by 
 $\tilde{C}^+_m$ which favors hopping and pair creation across small distances $m$.

\subsection{Reduction to the Jaynes-Cummings limit}
When we reduce the strength $g$ of the Rabi coupling sufficiently to reach the limit $g \ll \omega, \epsilon$ typical of the Jaynes-Cummings model, we can apply the RWA and drop counter-rotating terms. 
It is instructive to consider how, in this case, the effective Hamiltonian (\ref{Rabieffective}) reduces to the dispersive regime of the  Jaynes-Cummings lattice. 

Neglecting counter-rotating terms, the onsite interaction simplifies to the Jaynes-Cummings interaction
\be\label{JCint}
V=g \sum_j \left( a_j \sigma^+_j  +  \mathrm{H.c.} \right).
\ee
Neglecting all counter-rotating terms in an analogous derivation of the dispersive Hamiltonian, we only obtain contributions from paths {\scshape i} and {\scshape ii}.
Note that, generally, smaller energy differences between the intermediate and initial/final levels in figure \ref{ladders1} lead to larger effective coupling.  Further, $\Lambda$ and $V$-shaped paths have larger effective coupling due to a constructive sum of the two inverse energy differences.
Finally, the energy difference between the initial and final states for each path, when compared to the magnitude of the effective coupling,  determines whether or not a path contributes within the RWA.

The resulting effective Hamiltonian describing the Jaynes-Cummings lattice in the dispersive regime reads
\begin{eqnarray}
\nonumber  \tilde{H}_\mathrm{eff} =H_0+  g^2 \sum_{j\not=j'} C_{j-j'} \sigma_{j'}^+ \sigma_j^-     +      \frac{g^2}{2} \sum_{j\not=j'}C_{j-j'} a^\dag_j a_{j'}( \sigma^z_j + \sigma^z_{j'})  \\
\qquad\qquad   +  \frac{g^2}{2 } C_0 \sum_{j}( 2 a^\dag_j a_{j}+1 ) \sigma^z_j 
\label{JCeffective}           
 \end{eqnarray}
where we have again dropped a global constant. We define and approximate the involved coupling constants $C_m$ by
\be\label{hoppingfunction}
C_m =\frac{1}{N}\sum_k \frac{1 }{\Delta_k}e^{imk}=\frac{1}{N \Delta}\sum_k \frac{ e^{imk}}{1-\frac{2t}{\Delta} \cos{k}}  \approx \frac{1}{\Delta} \left(\frac{t}{\Delta}\right)^m, 
\ee
where we assume $0 \le m \le N/2$. (Outside this range, the same substitution $m\to|m\mathrm{\,mod\,}N|$ applies.)
Note that the difference between $\tilde{C}^+_m$ and $\tilde{C}^-_m$  disappears once the counter-rotating term $\sim 1/\Sigma_k$ is dropped.

The photon-mediated qubit-qubit interaction captured by the second term on the right-hand side of equation (\ref{JCeffective}) now has the typical flip-flop (XY) form reminiscent of the well-known ``quantum bus" interaction in the context of multiple qubits in a single resonator \cite{Majer2007}.
According to the form of the coupling constants $C_m$, this interaction is again short-range and decreases exponentially with the distance between lattice sites. Mediation of this interaction requires a virtual photon to hop across $m$ lattice sites, thus explaining the factor $(t/\Delta)^m$ in $C_m$ responsible for the short-range nature.  The third term describes the same conditional photon hopping discussed for the Rabi lattice above.  The fourth and final term combines the AC Stark and Lamb shifts of the on-site qubit-photon system, in agreement with the results in Ref.~\cite{Blais2004} when using the weak-coupling approximation $C_0\approx 1/\Delta$.

\section{Physics of the Jaynes-Cummings and Rabi lattice model in the dispersive regime \label{sec:phrab}}
Several previous studies of the single-site Rabi and Rabi lattice models have shown that interesting ground-state and steady-state properties emerge  in the ultra-strong coupling regime \cite{Peropadre2010,Nataf2010a,Ridolfo2012, Schiro2012, Hwang:2013kk}. As a particular example, the ground state of the Rabi lattice is believed to undergo a quantum phase transition involving symmetry breaking of the $Z_2$ parity in the ultra-strong coupling regime when qubit and photons are on resonance \cite{Schiro2012}. Here, we report that also the off-resonant, dispersive Rabi regime shows interesting ground-state properties when either the qubit or the photon frequency  is comparable to the interaction strength $g$.  Depending on the sign of the detuning $\Delta=\epsilon-\omega$, we distinguish between the dispersive Rabi regime with negative and positive detuning, respectively.

In the following three subsections, we discuss these two regimes along with the dispersive regime of the Jaynes-Cummings lattice. The most interesting aspect of the dispersive Jaynes-Cummings regime is the effective qubit-qubit interaction of  XY-type and the possibility of next-nearest-neighbor frustration. In the negative-detuning dispersive Rabi regime, this interaction turns into an effective transverse-Ising model  \cite{Schiro2012, Hwang:2013kk},  which predicts the same kind of phase transition as in Ref.\ \cite{Schiro2012}. The effective Hamiltonian we obtain includes additional non-nearest-neighbor Ising-type interactions, which were not considered in Ref.\ \cite{Hwang:2013kk}.  In the positive-detuning dispersive Rabi regime, we obtain an effective photonic Hamiltonian which shows  interesting one-mode and two-mode squeezing in its ground state.

\subsection{Qubit-qubit interaction in the dispersive Jaynes-Cummings regime}
In the Jaynes-Cummings regime, the interaction described by equation (\ref{JCint}) only swaps qubit and resonator excitations. As a consequence, the Hamiltonian has a $U(1)$ symmetry and the total number of excitations $N_\mathrm{tot}=\sum_j(a^{\dag}_j a_j  +  \sigma^+_j \sigma^-_j)$ is conserved.   In the dispersive Jaynes-Cummings regime, the inter-conversion of qubit and photon excitations is suppressed. By switching to a dressed-state basis,  the Schrieffer-Wolff transformation eliminates this interaction term. In that dressed-state basis, the total numbers of photon excitations, $N_\mathrm{ph}=\sum_j  a^{\dag}_j a_j$, and of qubit excitations, $N_\mathrm{qu}=\sum_j  \sigma^+_j \sigma^-_j$, are conserved separately and the  effective Hamiltonian, equation (\ref{JCeffective}),  hence possesses a $U(1) \times U(1)$ symmetry.  In other words, the effective Hamiltonian separates into blocks with fixed $N_\mathrm{ph}$ and $N_\mathrm{qu}$, thus greatly simplifying the numerical diagonalization.

The interaction terms emerging in the second-order treatment of the Jaynes-Cummings lattice are ordinary AC Stark shifts, conditional photon hopping terms, and qubit-qubit interaction of flip-flop (XY) type, see equation \eqref{JCeffective}.  We now focus on the qubit-qubit interaction. As usual, the effective flip-flop interaction can be rewritten as an XY interaction between (pseudo) spins,
\be\label{JCqq}
\tilde{H}_\mathrm{qubit-qubit}=g^2\sum_{j\not=j'} C_{j-j'} \sigma_{j'}^+ \sigma_j^- =  {\sum_{j>j'}} J_{j-j'} (\sigma_{j'}^x \sigma_j^x + \sigma_{j'}^y\sigma_j^y),     
\ee
with an interaction strength given by $J_m$.  Keeping the leading order term in $t/\Delta$, we can approximate the interaction strength as 
\be
J_m= \frac{g^2}{2 } C_m\approx \frac{g^2 }{2\Delta}\left(\frac{t}{\Delta}\right)^m.
\ee

An interesting fact to note is that $J_m$ can be tuned with the detuning $\Delta$ and the photon hopping strength $t$.  It is thus conceivable to engineer both ``ferromagnetic" (FM) or ``antiferromagnetic" (AF) qubit-qubit interactions, including the possibility of terms leading to non-nearest-neighbor frustration.  The signs for $J_1$ (nearest-neighbor), $J_2$ (next-nearest-neighbor), and the presence or absence of frustration is  summarized in Table \ref{summurytable} for the configurations that can occur. 

It is well known that
 the 1D antiferromagnetic $J_1$-$J_2$ XY and Heisenberg models  show a phase transition related to frustration and spontaneous dimerization \cite{Majumdar:1969iu,Haldane:1982el, Nishimori:1987iv,  Nomura:1993jz, Nomura:1994vf}.   According to Refs.~\cite{Nomura:1993jz, Nomura:1994vf}, the critical point for the $J_1$-$J_2$ XY model is given by $J_2/J_1 = 0.32$, which could indeed be accessible with the dispersive Jaynes-Cummings lattice where $J_2/J_1=t/ \Delta$.  [Recall: the necessary condition $\abs{t/\Delta}<1/2$ mentioned subsequent to equation \eqref{cplf} is indeed weaker and compatible with this value.] The frustration physics for the Jaynes-Cummings lattice, however, is more intricate: next-nearest neighbor frustration only occurs in the  \emph{positive}-detuning regime where the participation of photons is unavoidable.    In order to determine the fate of the phase transition, one thus needs to investigate the relevance of the photon terms in equation \eqref{Rabieffective} for the infinite chain near criticality, which is beyond the scope of this paper.

\begin{table}
\caption{\label{summurytable} Summary of the effective spin-spin interactions in the dispersive Jaynes-Cummings lattice, equations \eqref{JCeffective} and \eqref{JCqq}. Analogous statements hold for the dispersive Rabi lattice, equations \eqref{Rabieffective} and \eqref{Rabieffective2}, where $J_m$ should be replaced by $\tilde{J}_m$.}
\footnotesize
\begin{tabular*}{\textwidth}{@{}@{}l*{15}{@{\extracolsep{0pt plus12pt}}l}}
\br
detuning & hopping & nearest neighbor & next-nearest neighbor & frustration? \\
\mr
$\Delta>0$ & $t>0$ & $J_1>0$ (AF)   & $J_2>0$ (AF) & Yes\\
$\Delta>0$ & $t<0$ & $J_1<0$ (FM)   & $J_2>0$ (AF)&  Yes\\
$\Delta<0$ & $t>0$ & $J_1>0$ (AF)   & $J_2<0$ (FM)&  No\\
$\Delta<0$ & $t<0$ & $J_1>0$ (FM)   & $J_2<0$ (FM)&  No\\
\br
\end{tabular*}
\end{table}

\subsection{Dispersive Rabi regime for negative detuning \,\,($g \sim \epsilon \ll \omega $)}
In the case of negative detuning,  the qubit frequency $\epsilon$ is small compared to the photon frequency $\omega$ and we may perform a series expansion in the small parameter $\epsilon/\omega\ll1$.
Referring to equation \eqref{approxcoupling},  we find that the resulting coupling constants scale as
\be\fl\label{coefficients}
\tilde{C}^+_m = -\frac{2(m+1)}{\omega}\frac{\epsilon}{\omega}\left(\frac{-t}{\omega}\right)^m + \mathcal{O}(\epsilon^3/\omega^3), \qquad 
\tilde{C}^-_m =  -\frac{2}{\omega}\left(\frac{-t}{\omega}\right)^m + \mathcal{O}(\epsilon^2/\omega^2),
\ee
from which we infer that $\tilde{C}^+_m\ll \tilde{C}^-_m$ as long as $m$ is sufficiently small.

By inspection, we note that there are only two terms in the effective Hamiltonian \eqref{Rabieffective} which do not conserve photon number. These correspond to photon pair creation contributions with individual strengths set by $g^2 \tilde{C}^+_m$ with $m\ge 1$.  For large photon energies $\omega\gg\epsilon,\,t\,\,g$, it is clear that the inequality
\be
\omega \gg g^2 \tilde{C}^+_m \sim \epsilon \bigg(\frac{g}{\omega}\bigg)^2\left(\frac{t}{\omega}\right)^m
\ee
is satisfied automatically. Consequently, photon pair creation is strongly suppressed. Therefore, the resulting ground state is expected to be essentially free of photons in the dressed-state basis. 
These arguments lead us to the following dispersive Rabi model at negative detuning, valid in the $N_\mathrm{ph}=0$ manifold:

\begin{eqnarray}
 \tilde{H}_\mathrm{eff}\bigg|_{N_\mathrm{ph}=0} \approx  \frac{\epsilon+\tilde{K}_0}{2}\sum_j \sigma^z_j
+\sum_{j> j'} \tilde{J}_{j - j'} \sigma^x_{j}  \sigma^x_{j'} 
\label{Rabieffective2}
\end{eqnarray}
Here, the Ising coupling and Lamb shift parameter are given by 
\be
\tilde{J}_m=g^2 \tilde{C}^-_m \approx -\frac{2g^2}{\omega} \left(\frac{-t}{\omega}\right)^m
\quad\textrm{and}\quad \tilde{K}_0=g^2 \tilde{C}^+_0 \approx -2\frac{g^2\epsilon}{\omega^2}.
\ee
As before, the interaction strength $\tilde{J}_m$ decreases exponentially with the lattice site distance $m$. The relevant signs of the coupling parameters
 can be inferred from Table \ref{summurytable} (substituting $J_m\to \tilde{J}_m$). The renormalized qubit frequency $\epsilon'=\epsilon+\tilde{K}_0$ plays the role of the effective ``magnetic field" (up to a factor of 2), and tends to align the pseudo spin in negative $z$ direction.   However, the counter-rotating terms of the $\sigma^x_j\sigma^x_{j'}$ interaction compete with this tendency by tilting the pseudo spin towards the $xy$-plane.

For nearest-neighbor coupling only (i.e., $\tilde{J}_m\approx 0$ for $m>1$), we obtain a simple transverse-Ising model, which is exactly solvable via Jordan-Wigner transformation \cite{Sachdev1999}.    
In the thermodynamic limit (infinite chain), one recovers the usual quantum phase transition between a paramagnetic and a ferromagnetic phase. The transition occurs at $\tilde{J}_1=\epsilon/2$, which here implies a critical coupling of $g^\star = \frac{\omega}{2}\sqrt{\frac{\epsilon}{t}} $. (Corrections from weak non-nearest-neighbor interaction are expected to slightly shift this transition point.)
Below $g^\star$, the system is in the ``paramagnetic" (PM) phase -- with the ground state approximately given by the trivial vacuum state $\ket{\tilde{g}}_\mathrm{PM}   \approx  \ket{0}_\mathrm{ph}\otimes \ket{\!\!\downarrow\downarrow\downarrow\cdots\downarrow}$, where $\ket{0}_\mathrm{ph}$ denotes the photon vacuum and $\ket{\!\!\downarrow}$ the low-energy $\sigma^z$ eigenstate of each qubit. Above $g^\star$, the system is in the ``ferromagnetic" (FM) phase (assuming $t>0$). The two symmetry-broken ground-state wavefunctions of the system are  approximately given by $\ket{\tilde{g}_R}_\mathrm{FM} \approx  \ket{0}_\mathrm{ph}\otimes\ket{\!\!\rightarrow \rightarrow \rightarrow \cdots \rightarrow}$
and $\ket{\tilde{g}_L}_\mathrm{FM}  \approx \ket{0}_\mathrm{ph}\otimes\ket{\!\!\leftarrow \leftarrow \leftarrow \cdots \leftarrow}$, where $\ket{\!\!\rightarrow}$ and $\ket{\!\!\leftarrow}$ are the two $\sigma^x$ eigenstates.

The latter states, marked with ``$\tilde{\phantom{g}}$", denote eigenstates of the \emph{effective} Hamiltonian.
To obtain the eigenstates of the \emph{original} Hamiltonian, we perform the inverse Schrieffer-Wolff transformation  $\ket{g} = e^{-iS_1} \ket{\tilde{g}}$ with the previously used generator from equation \eqref{generatorRabi}. We illustrate the inverse transformation for the weak-hopping limit, $t \rightarrow 0$. In this case, the generator reduces to
\be\label{generatorRabionsite}
S_1 \approx i \sum_j \left[ \frac{g}{\Delta}(a_j^\dag \sigma^-_j -\mathrm{H.c.})   +  \frac{g}{\Sigma}(a_j \sigma^-_j -\mathrm{H.c.})   \right],
\ee 
and the inverse transformation decouples into mere onsite terms. Further approximating  $\Delta \approx -\omega$ and $\Sigma \approx \omega$, we obtain
\be\label{inverseapproximation}
\ket{g} = e^{-iS_1} \ket{\tilde{g}} \approx  \prod_j  \exp\bigg[ -\frac{g}{\omega}(a_j^\dag-a_j)\sigma^x_j\bigg]\ket{\tilde{g}}.
\ee 
For the first of the two ``FM" ground states, we evaluate
\be\label{inverseFM}
\ket{g_R}_\mathrm{FM} \approx  \prod_j   \exp\bigg[ -\frac{g}{\omega}(a_j^\dag-a_j)\sigma^x_j\bigg] \ket{0}_\mathrm{ph}\otimes\ket{\!\!\rightarrow \rightarrow \rightarrow \cdots \rightarrow}.
\ee
Since each qubit is in a $\sigma^x$ eigenstate, we recognize the remaining operator acting on the photon vacuum as a displacement operator $D(\mp \frac{g}{\omega})$ \cite{Gerry-Knight}, producing a coherent photon state on each site: 
\be\fl\label{approximateFM}
\ket{g_R}_\mathrm{FM} \approx  \prod_j \ket{\alpha_j=-{\textstyle\frac{g}{\omega}}}_j \otimes \ket{\!\!\rightarrow}_j \qquad \textrm{and}  \qquad \ket{g_L}_\mathrm{FM} \approx  \prod_j \ket{\alpha_j={\textstyle\frac{g}{\omega}}}_j \otimes \ket{\!\!\leftarrow}_j.
\ee
In a similar way,  we find 
\be\label{approximatePM}
\ket{g}_\mathrm{PM}  \approx  \prod_j \bigg[\ket{\alpha_j=-{\textstyle\frac{g}{\omega}}}_j \otimes \ket{\!\!\rightarrow}_j\,\,-\,\, \ket{\alpha_j={\textstyle\frac{g}{\omega}}}_j \otimes \ket{\!\!\leftarrow}_j \bigg].
\ee
for the paramagnetic ground state in the $t\to0$ limit. We note that these results are consistent with those obtained by a different method in Ref.\ \cite{Hwang:2013kk}.
In the general case, and particularly for quantitative comparison, the inverse transformation must be carried out for arbitrary hopping strength $t$, leading to further corrections to equations \eqref{approximateFM} and \eqref{approximatePM}.
We will properly account for this in our discussion in section \ref{sec4}.

Finally, we note that for the $N_\mathrm{ph} > 0$ manifolds, the situation is slightly more complicated.  According to equation \eqref{coefficients}, when keeping terms with coefficient $\tilde{C}^-_1$, we can ignore all  terms with coefficients $\tilde{C}^+_m$ except for those with $\tilde{C}^+_0$, which give rise to the Lamb shifts and the AC Stark shifts.  Thus, our effective Hamiltonian turns into a transverse-Ising model and a photon tight-binding model, with the only coupling between the two being the AC Stark shifts term, namely

\begin{eqnarray}\fl
\nonumber  \tilde{H}_\mathrm{eff} \approx   \frac{\tilde{K}_0 + \epsilon}{2}\sum_j \sigma^z_j
+\sum_{j} \tilde{J}_{1} \sigma^x_{j}  \sigma^x_{j+1}  + \omega  \sum_j  a^\dag_j a_j +t\sum_{\langle j,j' \rangle}\left(a_j^\dag a_{j'}  + \mathrm{H.c.} \right)  \\
+  g^2 \tilde{C}^+_0 \sum_{j} a^\dag_j a_{j} \sigma^z_j. 
\label{decoupling}
\end{eqnarray}

\subsection{Dispersive Rabi regime for positive detuning \,\,($g \sim \omega \ll \epsilon $)}
In the dispersive Rabi regime with positive detuning,  the photon frequency $\omega$ is small compared to the qubit frequency $\epsilon$. Thus, we may Taylor-expand in $\omega/\epsilon$ and find that the coupling constants $\tilde{C}^{\pm}_m$ now depend on whether $m$ is even or odd in the following way:  
\begin{eqnarray}
\label{C+}\fl
\textrm{odd }m:\quad \tilde{C}^+_m  = \frac{2(m+1)}{\epsilon}\frac{\omega}{\epsilon}\left(\frac{t}{\epsilon}\right)^m + \mathcal{O}(\omega^{3}/\epsilon^3),\\
\label{C+2}\fl
\textrm{even }m:\quad   \tilde{C}^+_m  = \frac{2}{\epsilon}\left(\frac{t}{\epsilon} \right)^m + \mathcal{O}(\omega^2/\epsilon^2).
\end{eqnarray}
For $\tilde{C}^-_m$ exactly the same expressions apply, except for an interchange of the roles of `even' vs.\ `odd'. Thus, in addition to the overall decrease in coupling with increasing lattice site distance $m$, we see that $\tilde{C}^+_m$ is further suppressed for odd $m$, whereas $\tilde{C}^-_m$ is further suppressed for even $m$.

Following a line of argument analogous to the one we employed for negative detuning, we note that now the effective Ising coupling  is small compared to the qubit frequency, $\tilde{J}_m \ll \epsilon$. As a result, we may neglect the counter-rotating terms $\sigma^+_{j} \sigma^+_{j'}+\sigma^-_{j} \sigma^-_{j'}$ of the Ising interaction. The remaining qubit-qubit interaction is then of XY-type, and conserves the total number of qubit excitations $N_\mathrm{qu}$. Since we are here interested in  the ground state and low-lying states only, we can restrict our discussion to the $N_\mathrm{qu}=0$ subspace.  In this  subspace, the effective dispersive Hamiltonian consequently takes the form
\begin{eqnarray}\label{squeezingHamiltonian}
\tilde{H}_\mathrm{eff}\big|_{N_\mathrm{qu}=0} \approx   \omega  \sum_j  a^\dag_j a_j +   t\sum_ j \left(a_j^\dag a_{j+1}  + \mathrm{H.c.} \right) \\\nonumber
\qquad\qquad\qquad   -   \frac{g^2}{2} \sum_{j, j'} \tilde{C}^+_{j-j'} (a^\dag_{j}a_{j'}+a^\dag_{j}a^\dag_{j'} + \mathrm{H.c.}).
\end{eqnarray}
This corresponds to a  photon tight-binding model with additional on-site and off-site photon pair creation/annihilation as well as additional, second-order photon hopping terms. Comparison with the expressions for $\tilde{C}^+_m$ in equations \eqref{C+} and \eqref{C+2} shows that onsite terms dominate the second-order contributions. Nearest neighbor and next-nearest neighbor terms are suppressed by factors of $\omega t/\epsilon^2$ and $t^2/\epsilon^2$, respectively.

By rewriting the effective Hamiltonian \eqref{squeezingHamiltonian} in $k$ space, we obtain
\be\label{reciprocalsqueezing}
\tilde{H}_\mathrm{eff}\big|_{N_\mathrm{qu}=0} = \sum_k \left[\omega_k \mathsf{a}^\dag_k \mathsf{a}_k + {\textstyle \frac{1}{2}}\delta_k  ( \mathsf{a}_k \mathsf{a}_{-k} + \mathsf{a}^\dag_k \mathsf{a}^\dag_{-k} )  \right].
\ee
The photon dispersion $\omega_k$ and photon pairing amplitude $\delta_k$ can be expressed as a Fourier series with coefficients determined by $\tilde{C}^+_m$.  We approximate them by truncating the series at $\tilde{C}^+_0$ and neglecting higher-order hopping and pairing.  That way, we find 
\be
\omega_k \approx  \omega+ 2t \cos{k} - 2g^2/\epsilon,   \qquad   \delta_k \approx -2g^2/\epsilon.
\ee
Here, the pairing amplitude has become $k$-independent since only on-site pairing is taken into account.   

We solve this Hamiltonian by performing the  Bogoliubov transformation
\be
b_k =u_k \mathsf{a}_k + v_k \mathsf{a}^\dag_{-k},  \qquad    b^\dag_{-k} =  v_k \mathsf{a}_k + u_k \mathsf{a}^\dag_{-k},
\ee 
where the coefficients $u_k$ and $v_k$ can be chosen real-valued and must satisfy $u_k=u_{-k}, v_k=v_{-k}$ and  $u_k^2 - v_k^2 =1$, such that canonical bosonic commutation relations hold for the new operators. As usual, we ensure these conditions by expressing the coefficients in the form $u_k=\cosh{r_k}, \ v_k=\sinh{r_k}$. Defining the remaining $r_k$ parameter via $\tanh{(2r_k)}=\frac{\delta_k}{\omega_k}$, the effective Hamiltonian is rendered diagonal, i.e., $\tilde{H}_\mathrm{eff}\big|_{N_\mathrm{qu}=0}  = \sum_k E_k b^\dag_k b_k$,
and the spectrum is given by
\be\label{spectrum}
E_k = \sqrt{\omega_k^2 - \delta_k^2} \approx \sqrt{\left(\omega+2t\cos{k} - 2g^2/\epsilon\right)^2 - 4g^4/\epsilon^2}.
\ee

Next, we show that the ground state (i.e., the `vacuum' of Bogoliubov excitations)  corresponds to a squeezed vacuum state of photons.  To see this, we recall the two-mode squeezing operators $\mathcal{S}_2(\xi_k)=\exp[\xi^*_k \mathsf{a}_k \mathsf{a}_{-k} - \xi_k  \mathsf{a}^\dag_k \mathsf{a}^\dag_{-k}]$ where $\xi_k=r_k  e^{i \varphi_k}$ is the squeezing parameter \cite{Walls1995,Gerry-Knight}. Note that  the Bogoliubov transformation is then equivalent to a two-mode squeezing transformation according to
\be
b_k  =  \mathcal{S}_2(\xi_k) \mathsf{a}_k \mathcal{S}^\dag_2(\xi_k) = \mathsf{a}_k  \cosh{r_k}  +e^{i \varphi_k} \mathsf{a}^\dag_{-k}  \sinh{r_k} ,
\ee
\be
b_{-k}  =  \mathcal{S}_2(\xi_k) \mathsf{a}_{-k} \mathcal{S}^\dag_2(\xi_k) = \mathsf{a}_{-k}  \cosh{r_k}  +e^{i \varphi_k} \mathsf{a}^\dag_k  \sinh{r_k}. 
\ee
Comparing to the results from the Bogoliubov transformation above,  we find that the squeezing parameters are given by $ \varphi_k=0$ and
 \be\label{squeezeparameter}
r_k=\frac{1}{2}\tanh^{-1}(\delta_k/\omega_k) \approx \frac{1}{2}\tanh^{-1}\left(\frac{-2g^2/\epsilon}{\omega+ 2t \cos{k} -2g^2/\epsilon}\right).
 \ee  
For the two special cases of $k=0$ or $k=\pi$ (center and edge of the first Brillouin zone), one obtains one-mode squeezing instead of two-mode squeezing. 
The ground state of the Rabi lattice in this regime can hence be expressed as a squeezed vacuum state,
\be\label{svs}
\ket{\tilde{g}}= \prod_{k \ge 0} \mathcal{S}_2(r_k) \ket{0} = \prod_{k \ge 0} \exp\bigg[r_k \mathsf{a}_k \mathsf{a}_{-k}-r_k \mathsf{a}^\dag_k \mathsf{a}^\dag_{-k} \bigg] \,\ket{0}.
\ee
involving entangled dressed photon pairs with opposite quasi-momenta.\footnote{Caveat: the photon pairs mentioned here are indeed \emph{dressed} photon pairs. To assess the situation in the basis of the original Hamiltonian, the ground state $\ket{g}=e^{-iS_1}\ket{\tilde{g}}$ should be calculated, and we will do so in section \ref{sec4}.}

It is useful to point out that equations \eqref{spectrum} and  \eqref{squeezeparameter} also reveal the necessary breakdown of perturbation theory when $g$ exceeds a critical value $g_c$. Specifically, when $\abs{\delta_k}>\omega_k$, the squeezing parameter is ill-defined and $E_k$ becomes imaginary. The resulting critical value
\be\label{breakdown}
g_c = \frac{1}{2}\min_k \sqrt{\epsilon(\omega-2t\cos{k})} = \frac{1}{2}\sqrt{\epsilon(\omega-2|t|)}
\ee
thus marks the maximum possible value for the convergence radius of the perturbative expansion.  In the context of a single Rabi site, the critical coupling strength $g_c=\frac{1}{2}\sqrt{\epsilon \omega}$ has previously been derived in Ref.~\cite{Ashhab2010}. We note that for positive detuning, $g\ll g_c$ is a more restrictive condition than the condition $g\ll\Delta_\mathrm{min}$. The inequality $g\ll g_c$ thus replaces equation \eqref{dispersivecondition} as the necessary condition for the dispersive regime at positive detuning. As $g$ is increased beyond $g_c$, perturbation theory is no longer valid. Entering this quasi-resonant regime of the Rabi lattice,  it is plausible that the system will undergo the same type of phase transition with $Z_2$ symmetry breaking that was studied by Schiro et al.\ \cite{Schiro2012} for the case of exact resonance, $\epsilon=\omega$.

We illustrate the dressed-photon ground state by calculating several observables related to photon numbers and pairing amplitudes. The ground state expectation value for the photon number in mode $k$ is given by
\be\fl
\boket{\tilde{g}}{\mathsf{a}^\dag_k \mathsf{a}_k}{\tilde{g}}=\boket{0}{\mathcal{S}^\dag_2(r_k)\mathsf{a}^\dag_k \mathcal{S}_2(r_k)\mathcal{S}^\dag_2(r_k)\mathsf{a}_k \mathcal{S}_2(r_k)}{0} = \sinh^2(r_k).
\ee
Similarly, we find that the pair amplitude for dressed photons of opposite quasi-momenta is
\be
\boket{\tilde{g}}{\mathsf{a}_k \mathsf{a}_{-k}}{\tilde{g}}=\boket{\tilde{g}}{\mathsf{a}^\dag_k \mathsf{a}^\dag_{-k}}{\tilde{g}} = -\sinh{r_k} \cosh{r_k}.
\ee
No such pairing occurs for photons with identical quasi-momenta,
\be
\boket{\tilde{g}}{\mathsf{a}_k \mathsf{a}_{k}}{\tilde{g}}=\boket{\tilde{g}}{\mathsf{a}^\dag_k \mathsf{a}^\dag_k}{\tilde{g}} = 0, 
\ee
as long as the quasi-momenta $k$ and $-k$ are distinct (i.e., $k=0$ and $k=\pi$ are excluded). All these are the usual properties of two-mode squeezed states \cite{Walls1995}.

\section{Application to the Rabi dimer and comparison with results from exact diagonalization\label{sec4}}
In order to illustrate the utility of the effective dispersive Hamiltonian, and to demonstrate its validity by comparison with results from exact numerical diagonalization, we now turn to the specific example of the Rabi dimer, i.e., two coupled Rabi sites. We choose this example to make exact diagonalization of the full Rabi lattice Hamiltonian \eqref{Rlat} as tractable as possible, and emphasize that results obtained from the \emph{effective} Hamiltonian easily carry over to larger lattices, as evident from our equation \eqref{Rabieffective}. For verification of our approximations, we select representative observables, and calculate their expectation values by exact diagonalization of the original Rabi lattice Hamiltonian \eqref{Rlat}. We then compare these results with those obtained from our effective Hamiltonians  \eqref{decoupling} and \eqref{squeezingHamiltonian}, which are simplified approximations to Hamiltonian \eqref{Rabieffective} in the negative- and positive-detuning regimes respectively using the full expressions for the coupling constants $\tilde{C}^{\pm}_m$ from equation \eqref{cplf}.\footnote{An additional replacement $t \rightarrow t/2$ is performed to account for the fact that a two-site Rabi ring is equivalent to a dimer except for a factor of 2 in the hopping amplitude.}

\begin{figure}
\includegraphics[width=1.0\columnwidth]{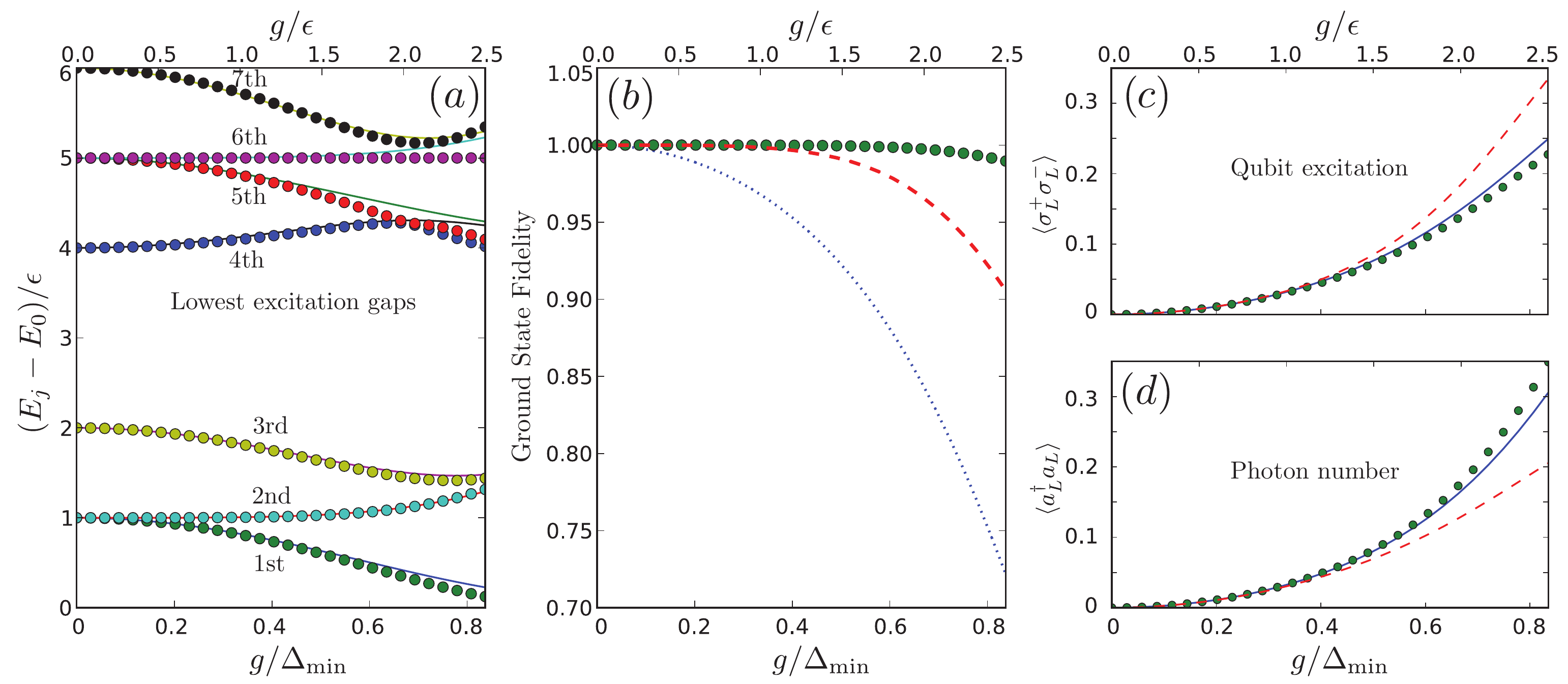}
\caption{Rabi dimer in the dispersive regime with negative detuning: comparison of representative observables between exact diagonalization and effective Hamiltonian. All panels show observables as a function of the Rabi coupling $g$.
 (a) Lowest excitation gaps $E_j - E_0$ between level $j$ and the ground state.  Solid curves show results from exact numerical diagonalization, circles mark results obtained from the effective Hamiltonian  \eqref{decoupling} .    (b) Fidelity of the ground state wavefunction with respect to the exact result. Green circles (full inverse SW transformation) and the red dashed curve (truncated inverse SW transformation) depict results from the effective Hamiltonian, differing only in the order of the inverse Schrieffer-Wolff transform (see main text). Blue dotted line: fidelity of the vacuum state for comparison. (c) and (d) Ground-state expectation values for number of qubit and photon excitations on the left Rabi site. The comparison shows exact results (blue solid curve), and results from the effective Hamiltonian with two different inverse transformation schemes (green circles and red dashed curve).  Over a wide range of $g$, the results from the effective Hamiltonian agree well with the exact results until $g$ approaches $\Delta_\mathrm{min}$, where the perturbation is known to break down.  (Parameters used: $\omega=5\epsilon, \  t=\epsilon$; photon cutoff per site: $n_c=12$.)}
\label{neg}
\end{figure}

\subsection{The dispersive regime with negative detuning}
Figure \ref{neg} shows the comparison for the dispersive regime with negative detuning, where example parameters have been chosen as
$\omega=5\epsilon$ and $ t=\epsilon$.  With this choice of $t$, the inequality $\abs{t/\Delta}<1$ holds and the dispersive condition \eqref{dispersivecondition} can be satisfied.   For the Rabi dimer with positive $t$,   equation \eqref{dispersivecondition} takes the simple form $\Delta_\mathrm{min}=\omega-|t|-\epsilon \gg g$ , where $\Delta_\mathrm{min}$ corresponds to the detuning of the antisymmetric photon mode. Perturbation theory and the effective Hamiltonian are expected to work reasonably well as long as $g/\Delta_\mathrm{min}$ is sufficiently small. This is indeed confirmed by figure \ref{neg}. The individual results from the four panels are as follows.

Panel (a) shows the lowest seven excitation gaps $E_j- E_0$ ($E_0$ being the ground-state energy and $E_j$ the $j$-th excited-state energy) as a function of the Rabi coupling strength $g$.  The lower three gaps correspond to states in the $N_\mathrm{ph}=0$ manifold, the remaining four to the $N_\mathrm{ph}=1$ manifold. The different curves correspond to exact numerical diagonalization on one hand, and results using the effective Hamiltonian  \eqref{decoupling} on the other hand.  We find very good agreement between approximate and exact results, up to values as high as $g/\Delta_\mathrm{min}\simeq 0.8$. As expected, lower gaps in manifolds with lower photon number match comparatively better. Perturbation theory must break down for $g/\Delta_\mathrm{min}\ge1$, as the qubit reaches quasi-resonance with the anti-symmetric photon mode at this point.  
Note that for small $g/\Delta_\mathrm{min}$ the 5th to 7th excitation gap differs from the 1st to 3rd excitation gap by a frequency $4\epsilon$, which corresponds to the photon frequency of the antisymmetric mode.  This recurrence illustrates the approximate decoupling of the transverse Ising and tight-binding model in the effective Hamiltonian \eqref{decoupling}.

Panel (b) shows the fidelity of the approximate ground-state wavefunctions, as obtained from the effective Hamiltonian \eqref{decoupling} [or, for the ground state,  equivalently to the transverse-Ising model, equation \eqref{Rabieffective2}]   and subsequent inverse Schrieffer-Wolff (SW) transformation $\ket{g} = e^{-iS_1} \ket{\tilde{g}}$.  The generator $S_1$ we choose here is the exact expression from Eq.~(\ref{generatorRabi}) and we do not take the weak hopping ($t \rightarrow 0$) limit.   For careful comparison,  we perform the inverse transformation with two different schemes:  (1) we apply the transformation directly in its full exponential form $ e^{-iS_1}$ (full inverse SW transformation), and  (2) we apply the transformation but keep only terms up to second order in $g$, namely $e^{-iS_1}=1-iS_1 +\frac{1}{2} (-iS_1)^2 + \mathcal{O}(g^3)$ (truncated inverse SW transformation). (This truncation would be used for consistently keeping only terms up to second order.)
For comparison, we also show the fidelity of the trivial vacuum state $\ket{0}_\mathrm{ph}\otimes \ket{\!\!\downarrow\downarrow\downarrow\cdots\downarrow}$.  We observe that for the $g/\Delta_\mathrm{min} \rightarrow 0$ limit, all three fidelities are similar and are very close to a 100\% value.  For large $g$, the vacuum fidelity decreases significantly, as expected in the ultra-strong coupling regime. Using the full inverse SW transform, the fidelity of the approximate state is very good, exceeding a  99\% value over the full range shown and gives better results as obtained with the truncated transform.

In panels (c) and (d)  we show plots for the ground-state expectation values of the qubit excitation $\langle \sigma^+_L  \sigma^-_L   \rangle$ and the photon number $\langle a_L^\dag a_L \rangle$ (both measured on one of the two Rabi sites). The agreement between exact and approximate results  is excellent over the entire range of $g$ in the plot. As before, results are slightly better when using the full inverse SW transform instead of the truncated version. 
It is interesting to note that the number of qubit excitations on each site rises to about $0.25$ around $g/\epsilon=2.5$, indicating that the qubits are tilting up towards  the $xy$-plane.  The tilting  is mainly induced by the transverse Ising interaction between qubits,  namely the $\sigma^x_L \sigma^x_R$ term.  We have confirmed numerically that this tilting is negligible for a single-site Rabi system, in which no transverse Ising interaction is present. 
.

 Although the effective Hamiltonian \eqref{Rabieffective2} produces a ground state with zero dressed photons in the negative detuning regime, the undressing from the inverse Schrieffer-Wolff transformation induces small corrections.  This effect can be motivated from  equations \eqref{inverseapproximation}--\eqref{approximateFM}, which indicate that the $\sigma^x$ qubit eigenstate is always accompanied by a photon coherent state on the same site, namely  $ \ket{\alpha_j={-\textstyle\frac{g}{\omega}}}_j \otimes\ket{\!\!\rightarrow}_j$.   For the finite-size Rabi dimer, the transverse-Ising interaction cannot give rise to an actual phase transition to an ordered spin state. Such a phase transition may, however, manifest in the thermodynamic limit (infinite lattice size) but is beyond the scope of our current paper.

\subsection{The dispersive regime with positive detuning}

\begin{figure}
\includegraphics[width=1.0\columnwidth]{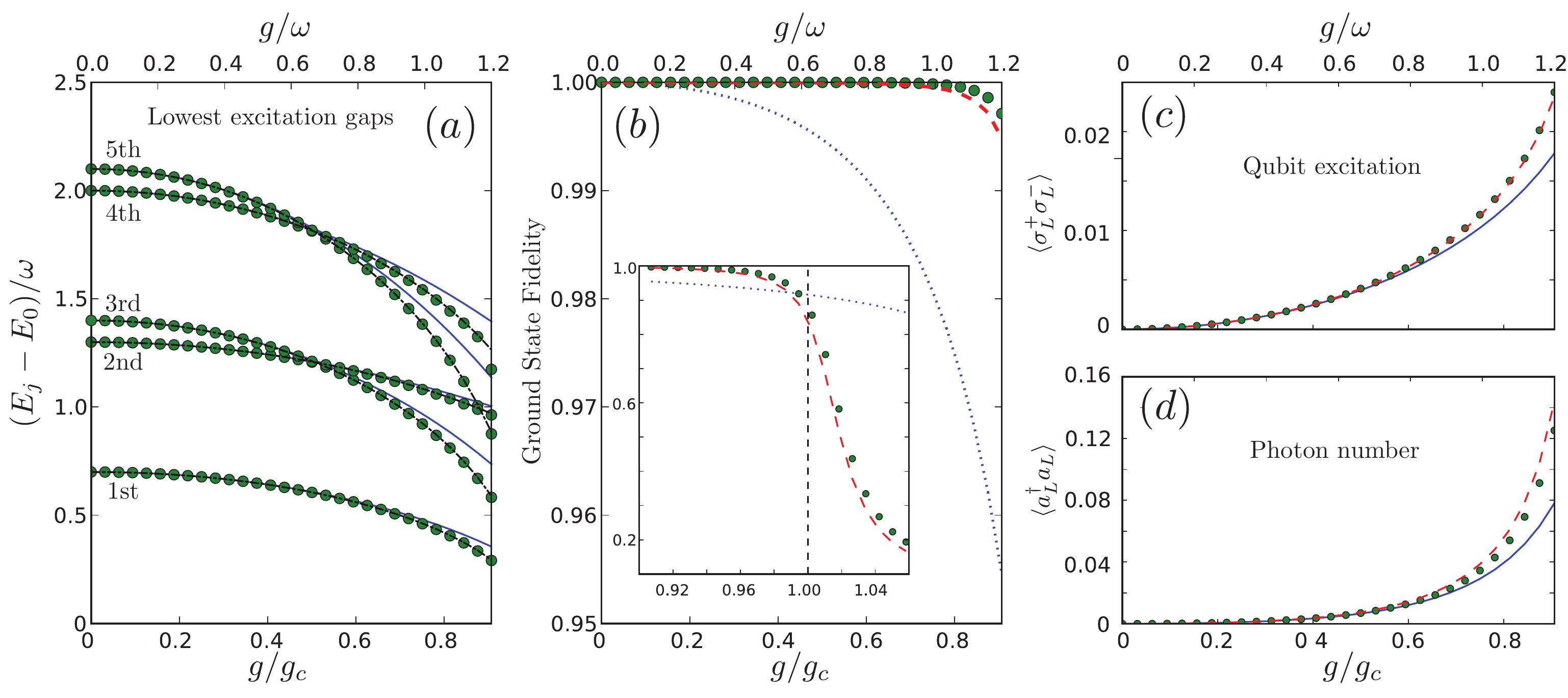}
\caption{
Rabi dimer in the dispersive regime with positive detuning: comparison of representative observables between exact diagonalization and effective Hamiltonian \eqref{squeezingHamiltonian}. All panels show observables as a function of the Rabi coupling $g$, and are arranged analogous to figure \ref{neg}:
 (a) Lowest excitation gaps $E_j - E_0$ between level $j$.
(b) Fidelity of the ground state wavefunction with respect to the exact result. The inset depicts the same fidelity plot but close to the critical coupling $g/g_c=1$. (c) and (d) show the ground-state expectation values for number of qubit and photon excitations on the left Rabi site. The legend follows figure \ref{neg}, i.e., exact diagonalization results: solid curves; results from effective Hamiltonian: circles (no truncation) and dashed curves (with truncation); fidelity of pure vacuum state: blue dotted curve; additional dot-dashed curves in panel (a):  analytical expressions of quasiparticle energies from equation \eqref{spectrum}.  Over the wide range of $g$, the results from the effective Hamiltonian agrees well with the exact results until $g$ approaches the critical value $g_c$. 
(Parameters: $\epsilon=10\omega, \  t=0.3\omega$; resulting critical coupling: $g_c=1.32\omega$; photon cutoff per site:  $n_c=12$.) }
\label{pos}
\end{figure}

We next turn to the dispersive regime of the Rabi dimer with positive detuning.
Figure ~\ref{pos} shows a comparison analogous to that presented in figure \ref{neg}, now  with model parameters fixed to $\epsilon=10\omega$ and $t=0.3\omega$.   Here,  we compare the original Rabi-lattice Hamiltonian \eqref{Rlat} to the effective Hamiltonian in the $N_\mathrm{qu}=0$ manifold, namely  equation \eqref{squeezingHamiltonian}.  We investigate the same set of observables as in the previous subsection and plot them as a function of $g$, here in units of the critical coupling strength $g_c=\frac{1}{2}\sqrt{\epsilon(\omega-|t|)}$ [here we have already carried out the $t\rightarrow t/2$ replacement relative to equation \eqref{breakdown}] which is the relevant quantity marking the breakdown of perturbation theory for positive detuning. The critical interaction strength for our choice of parameters is given by $g_c= 1.32\omega$.

The lowest five excitation gaps plotted in panel \ref{pos}(a) show very good agreement between exact and approximate results using the effective Hamiltonian over a wide coupling range, with best agreement for the lowest gap. Here,  solid lines and circles represent results from exact diagonalization and diagonalizing the low-lying effective Hamiltonian \eqref{squeezingHamiltonian} respectively,  while the dot-dashed lines represent results using the analytical expression equation \eqref{spectrum} for the quasiparticle energies (the quasimomentum $k=0$ and $k=\pi$ correspond to the symmetric and anti-symmetric photon mode respectively).    In the $g=0$ limit, original and Bogoliubov operators coincide, $b_k =\mathsf{a}_k$, and excitations correspond to photon Fock states in the two modes. Specifically, for our parameters the 1st, 3rd and 5th excitation gaps correspond to creation of 1, 2 and 3 photons in the anti-symmetric mode with frequency $\omega-t=0.7\epsilon$.  The 2nd excitation gap corresponds to creation of 1 photon in the symmetric mode, which here has frequency $\omega+t =1.3\epsilon$ etc..
For $g>0$, the Bogoliubov operators $b_k$ involve both photon annihilation ($\mathsf{a}_k$) and creation ($\mathsf{a}^\dag_k$) operators. The eigenstates are no longer pure Fock states and the excitation energies decrease as a function of $g$, as expected from equation~\eqref{spectrum}.

Panel \ref{pos}(b) shows the ground-state fidelities for our perturbative approximations as well as the trivial vacuum state.    For $g$ near $0$,   the ground state remains quite  close to the trivial vacuum state.   However,  as $g$ is further increased,  the expected squeezing of the ground state becomes more significant and the fidelity of the vacuum state drops significantly below the fidelities of our perturbative approximations,  which remain very close to 1 until $g$ approaches the critical value $g_c$.   The expected breakdown of the perturbative treatment close to $g=g_c$ is shown in the inset of panel (b).

Panels \ref{pos}(c) and (d) show the expected  photon number and qubit excitation on one of the two Rabi sites for the Rabi dimer ground state. The exact-diagonalization results based on equation \eqref{Rlat} and our perturbative results based on the effective Hamiltonian \eqref{squeezingHamiltonian} show good agreement in the relevant range of coupling strengths.  Note that, by contrast to the situation of negative detuning,  the photon number here exceeds the expected value $\langle\sigma^+_L\sigma^-_L\rangle$ of qubit excitation.  Since the ground state is in the $N_\mathrm{qu}=0$ manifold,  the creation of qubit excitations is merely due to the qubit-photon dressing and is recovered as a correction from the inverse Schrieffer-Wolff transformation.
We note that due to the qubit-photon dressing, an additional increase in the photon number arises from the qubit excitation, as dictated by the inverse Schrieffer-Wolff transformation.

\subsubsection{Visualization of squeezing and photon pairing in the Rabi dimer} 
\begin{figure}
\includegraphics[width=0.7\columnwidth]{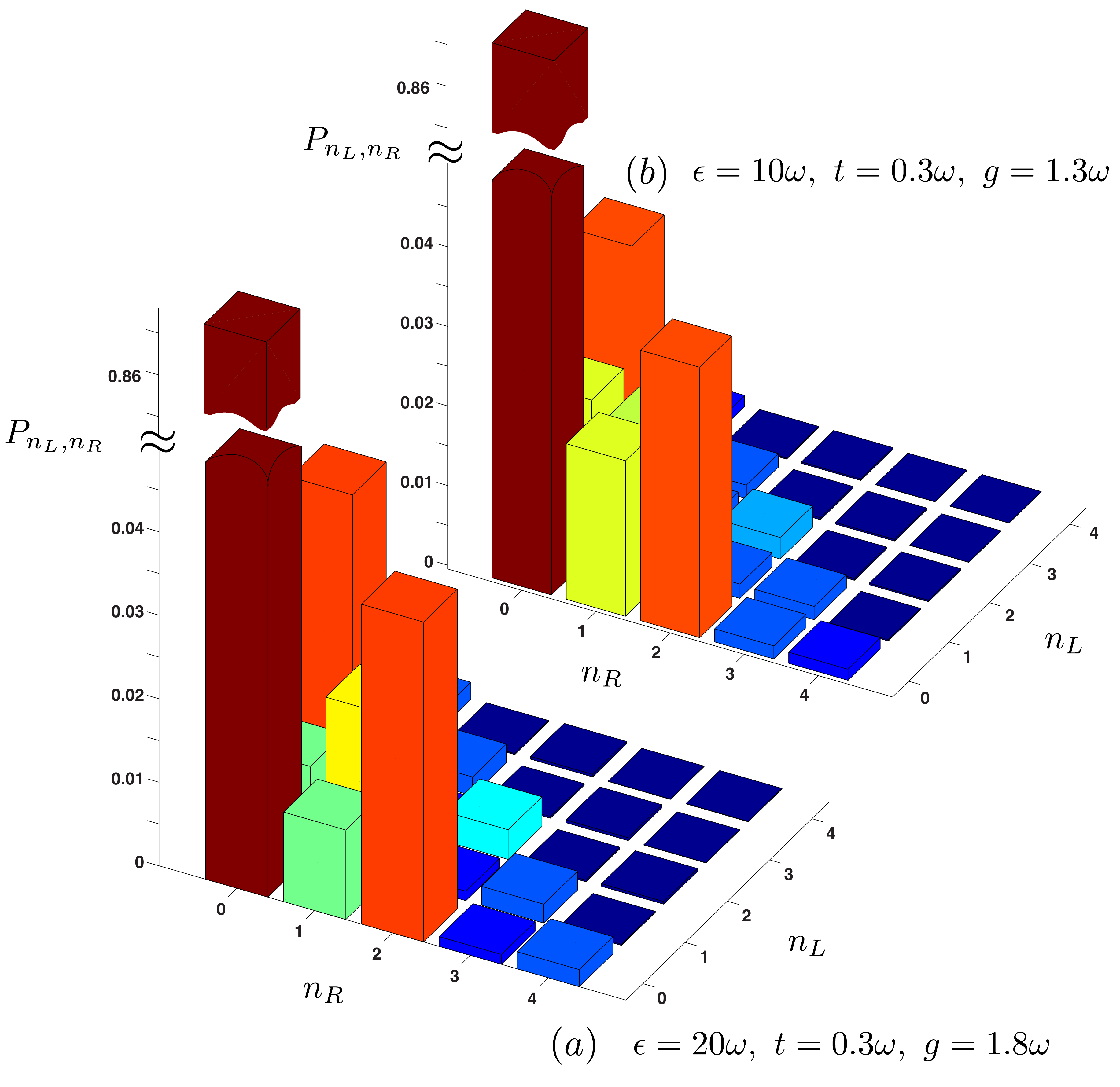}
\centering
\caption{Rabi dimer in the dispersive regime with positive detuning: joint-probability distribution $P_{n_L, n_R}$  of the ground state in the dimer Fock basis ($n_L, n_R$: photon number on the left/right site).  The results are obtained from exact diagonalization of the full Rabi-lattice Hamiltonian \eqref{Rlat}  with a photon cutoff $n_c=16$ on each site.  The two panels show the distributions for two different parameter sets, as specified in the figure.    Note that the distribution $P_{n_L, n_R}$ for even-number Fock states, $n_L+n_R=2N$,  is much larger than their neighboring odd-number Fock states, $n_L+n_R=2N \pm 1$, which evidently serves as a signature of photon pairing.    This pairing signature for parameter set (a) is more significant than the one for set (b), due to the larger detuning of set (a)  and hence smaller dressing effect which breaks photon parity.}
\label{jointdistribution}
\end{figure}
In order to elucidate the effect of the photon pairing or squeezing terms in  \eqref{squeezingHamiltonian}, we briefly discuss results for the Fock-state probability distribution and the Wigner function describing the Rabi dimer.

Figure~\ref{jointdistribution} shows the Fock-state probability distribution of the Rabi dimer ground state,
\be
P_{n_L,n_R}=|\mathrm{tr}_{\sigma_L,\sigma_R}\langle n_L,\sigma_L;n_R,\sigma_R|g\rangle|^2,
\ee
where $n_L,n_R$ denote the photon numbers on each Rabi site and we have traced out the qubit degrees of freedom. 
The distribution confirms that the pure vacuum state, even though dominant in the distribution, is not the true ground state, as expected for ultra-strong coupling. Expressed in terms of dressed photon states, equation \eqref{svs} predicts that the ground state should only involve even-number Fock states, $n_L+n_R=2N$, while the probability for odd-number Fock states should vanish. Due to the undressing by the inverse Schrieffer-Wolff transform, corrections to this simple picture emerge. As seen in the comparison of panels (a) and (b), these corrections become more significant as the detuning is decreased. In both cases, however, we clearly observe the fingerprint of photon pairing:   even-number Fock states with $n_L+n_R=2N$,  have higher probability than their neighboring odd-number Fock states with  $n_L+n_R=2N \pm 1$.   
By comparing $P_{2,0}$ and $P_{1,1}$, we also observe that onsite pair creation is more significant than offsite pair creation. This agrees well with the fact that  $\tilde{C}^+_0$ is larger than $\tilde{C}^+_1$, as we demonstrated in equation~\eqref{C+}.

\begin{figure}
\includegraphics[width=1\columnwidth]{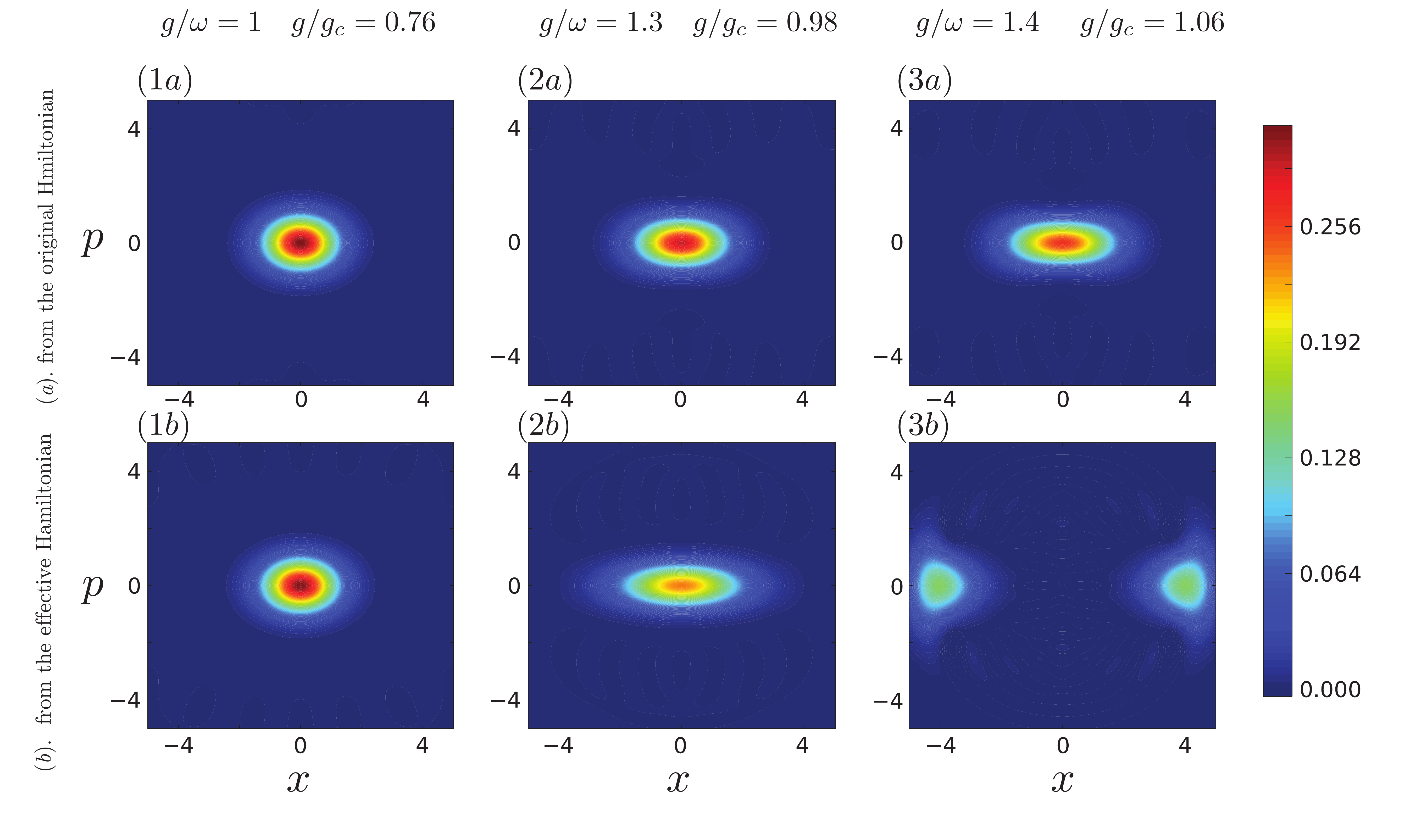}
\caption{Rabi dimer in the dispersive regime with positive detuning: Wigner functions of the reduced photon state on one Rabi site. For comparison, results from the original Rabi-lattice Hamiltonian and from the effective Hamiltonian are shown in row (a) and row (b), respectively.  Different columns show increasing coupling strengths $g$.  The shape of the Wigner distribution for $g/\omega = 1, 1.3$ ($g/g_c = 0.76, 0.98$) reveals vacuum squeezing of the ground state.   Results from the original and effective model agrees well as long as $g<g_c$.    Note, that the splitting of the Wigner distribution for the effective model at $g/ \omega=1.4$ $ (g/g_c=1.06)$ occurs earlier than the splitting of the exact Wigner function at approximately $g/g_c=1.16$ (not shown).
(Parameters: $\epsilon=10\omega$, $t=0.3\omega$; photon cutoff  $n_c = 16$ per site.)}
\label{wigner}
\end{figure}
An alternative way of visualizing the squeezing predicted by equations \eqref{squeezingHamiltonian} and \eqref{reciprocalsqueezing} is to calculate and plot the Wigner function for the reduced density matrix representing the photon state on one of the two Rabi sites. Given the Hilbert space structure of the dimer,  $\mathcal{H}_\mathrm{dimer}= \mathcal{H}^\mathrm{ph}_L \otimes \mathcal{H}^\mathrm{qu}_L \otimes  \mathcal{H}^\mathrm{ph}_R \otimes \mathcal{H}^\mathrm{qu}_R$, we can obtain the desired reduced density matrix directly from the calculated ground state,
\be
\rho = \mathrm{tr}_{n_R}\,\mathrm{tr}_{\sigma_L}\,\mathrm{tr}_{\sigma_R} \;|g\rangle\langle g|.
\ee
 Therefore,  we perform a partial trace of the ground state density matrix  over the Hilbert space except for the subspace $\mathcal{H}^\mathrm{ph}_L$, namely the photon Fock space on the left site.  The  Wigner function is then  obtained via $W(x,p)=2\pi^{-1}\mathrm{tr}\,[D(-\alpha)\rho\, D(\alpha)\,\mathcal{P}]$, where $D(\alpha)$ is the usual displacement operator, $\alpha=x+ip$ and $\mathcal{P}=\exp[i\pi a^\dag a]$ is the photon number parity operator \cite{Haroche2006}. 

The resulting Wigner function is plotted in figure~\ref{wigner} for several values of $g/\omega$, with the first row showing the result obtained from the full Rabi lattice Hamiltonian \eqref{Rlat} and the second row the corresponding result calculated from the dispersive Hamiltonian \eqref{Rabieffective}.
For $g=\omega$ ($g/g_c=0.76$),  the exact and approximate Wigner functions are in good agreement and show the expected squeezing.   Increasing the coupling further to  $g=1.3\omega$ ($g/g_c=0.98$), squeezing becomes more significant and deviations between approximate and exact result are visible as the breakdown point $g/g_c=1$ is approached.      For $g=1.4\omega$ ($g/g_c=1.06$), we exceed the critical coupling and the Wigner functions differ significantly, signaling the breakdown of perturbation theory.

Overall, our comparison between exact and approximate results for the Rabi dimer in the last two subsections nicely confirms the validity of the perturbative approach in the expected parameter regimes.

\section{Conclusion and Outlook\label{sec5}}
In summary, we have derived the effective Hamiltonian for the dispersive regime of the Rabi lattice by employing a Schrieffer-Wolff transformation  to second order in the Rabi interaction $\sim g$.    Our results generalize the well-established treatment of the dispersive limit for a single Rabi site to the case of the Rabi lattice, which has enjoyed substantial recent interest in the context of photon-based quantum simulation.  
The effective interaction terms emerging from our treatment include transverse-Ising interaction between qubits (XY interaction in the Jaynes-Cummings limit), photon pairing terms and conditional photon hopping terms. We have presented analytical expressions for the resulting coupling constants and demonstrated that they are short-range but not restricted to nearest-neighbor sites.

As necessary conditions for the validity of the dispersive regime of the Rabi lattice, we identified the inequalities 
\[
g\ll \mathrm{min}_k\, |\epsilon-\omega_k|=\omega-2|t|-\epsilon \quad \textrm{(negative detuning)}
\]
and
\[
g\ll \mathrm{min}_k\, \sqrt{\epsilon\omega_k}/2=\sqrt{\epsilon(\omega-2|t|)}/2 \quad \textrm{(positive detuning)}.
\]
For negative detuning, we found that the effective spin physics is given by a transverse-Ising model which includes interaction terms beyond nearest-neighbor spins. We showed how to recover the dressing of these spin states by photon coherent states via the inverse Schrieffer-Wolff transformation. For positive detuning, we studied the manifold of states without qubit excitations and discussed the effects of one-mode and two mode squeezing.

We  confirmed the validity of our effective model numerically and discussed in detail  the Rabi dimer as the simplest non-trivial example of a Rabi lattice model. 

Interesting future work should include extending the perturbative treatment to fourth order, where additional photon-photon interaction is expected due to self-Kerr and cross-Kerr terms.  An interesting open question is whether the phase transition discussed in Ref.~\cite{Schiro2012} can be accessed from within the dispersive Rabi regime with positive detuning when including such higher-order terms. Another interesting question concerns the fate of frustration induced phase transitions in the presence of spin-photon dressing. Finally,  consideration of the open-system aspect including dissipation and external driving forms an important theoretical challenge in the near future.

\ack
We thank Gianni Blatter, Andrew Houck, 
Hakan T\"ureci, and C.~Y.~Li for valuable discussions. This work was supported
by the Swiss National Science Foundation (SNF) and the NSF under
CAREER award PHY-1055993.

\section*{References}

\bibliographystyle{iopart-num}
\bibliography{library-merged}

\end{document}